\documentclass{emulateapj}
\newcommand{\alow}{$\alpha_{\mathrm{low}}$}
\newcommand{\amid}{$\alpha_{\mathrm{mid}}$}
\newcommand{\ahigh}{$\alpha_{\mathrm{high}}$}
\newcommand{\flow}{$R_{\mathrm{low}}$}
\newcommand{\ftherm}{$f_{\mathrm{th}}$}
\newcommand{\fthermL}{$f_{\mathrm{th,1.4}}$}

\def \eg{{\em e.g.}}

\def \ie{{\em i.e.}}
\def \via{{\em via}}

\shorttitle{Integrated Radio Continuum Spectra of Galaxies}
\shortauthors{Marvil et al.}

\begin{document}

\title{Integrated Radio Continuum Spectra of Galaxies}

\author{Joshua Marvil\altaffilmark{1}}
\affil{National Radio Astronomy Observatory\altaffilmark{2}, Socorro, NM 87801}
\affil{New Mexico Tech, Socorro, NM 87801}

\author{Frazer Owen}
\affil{National Radio Astronomy Observatory, Socorro, NM 87801}

\and

\author{Jean Eilek}
\affil{National Radio Astronomy Observatory, Socorro, NM 87801}
\affil{New Mexico Tech, Socorro, NM 87801}

\altaffiltext{1}{present address: CSIRO Astronomy and Space Science, 
PO Box 76, Epping, NSW 1710, Australia;  email: josh.marvil@csiro.au}
\altaffiltext{2}{The National Radio Astronomy Observatory is a facility of the National Science Foundation operated under cooperative agreement by Associated Universities, Inc.}

\begin{abstract}
We investigate the spectral shape of the total continuum radiation, between 74 MHz and 5 GHz (400 to 6 cm in wavelength), for a large sample of bright galaxies. We take advantage of the overlapping
survey coverage of the VLA Low-Frequency Sky Survey, the Westerbork Northern Sky Survey,  the NRAO VLA Sky Survey, and the Green Bank 6 cm survey to achieve significantly better resolution, sensitivity, and sample size compared to prior efforts of this nature. For our sample of 250 bright galaxies we measure a mean spectral index, $\alpha$, of -0.69 between 1.4 and 4.85 GHz, -0.55 between 325 MHz and 1.4 GHz, and -0.45 between 74 and 325 MHz, which amounts to a detection of curvature in the mean spectrum.  
The magnitude of this curvature is approximately $\Delta \alpha =  -0.2$ per logarithmic frequency decade when fit with a generalized function having constant curvature.  
No trend in low frequency spectral flattening versus galaxy inclination is evident in our data, suggesting that free-free absorption is not a satisfying explanation for the observed curvature.  The ratio of thermal to non-thermal emission is estimated by two independent methods, (1) using the IRAS far-IR fluxes, and (2) with the value of the total spectral index.   Method (1) results in a distribution of 1.4 GHz thermal fractions of 9\% $\pm$ 3\%, which is consistent with previous studies, while method (2) produces a mean 1.4 GHz thermal fraction of 51\% with dispersion 26\%. 
The highly implausible values produced by method (2) indicate that the sum of typical power-law thermal and non-thermal components is not a viable model for the total spectral index between 325 and 1.4 GHz. 
An investigation into relationships between spectral index, infrared-derived quantities and additional source properties reveals that galaxies with high radio luminosity  in our sample are found to have, on average, a flatter radio spectral index, 
and early types tend to have excess radio emission when compared to the radio-infrared ratio of later types.
Early types also have radio emission which is more compact than later type galaxies, as compared to the optical size of the galaxy. Despite these differences, no relation between spectral index and galaxy type is detected. 
\end{abstract}


\section{Introduction}\label{secInt}

\begin{figure}
\epsscale{1.20}
\plotone{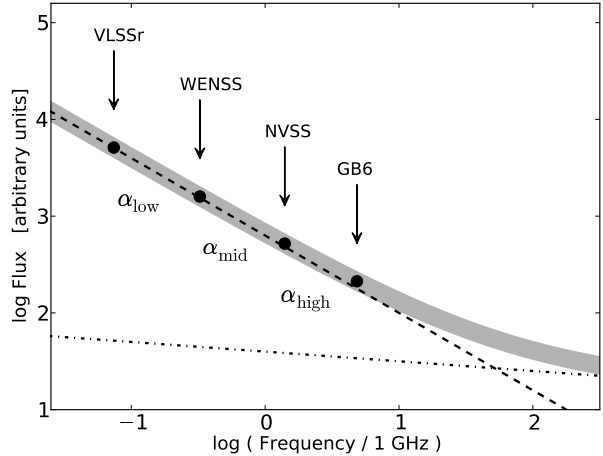}
\caption{Illustration of a generic radio spectrum with basic power-law model components:  steep spectrum synchrotron (dashed) with spectral index $\alpha_{NT} = -0.8$, flat-spectrum free-free (dash-dot) with spectral index $\alpha_{T} = -0.1$, and sum (shaded).  The locations of the radio survey data  are indicated with filled circles.  Spectral indices \alow, \amid, and \ahigh\ are calculated from the total radio spectrum between adjacent frequency intervals.}
\label{figToon1}
\end{figure}

Early radio continuum observations of bright galaxies established an approximately power-law relation between flux $S$ and frequency $\nu$, typically parameterized by the spectral index 
$\alpha$, defined here as the positive power-law exponent:  $S(\nu) \propto \nu^{\alpha}$. 
The spectral index is an important observational quantity since it can be interpreted as a measure of a source's physical properties and processes. 
The spectra below $\sim 50$ GHz from star forming galaxies are typically modeled as the sum of synchrotron emission, with non-thermal spectral index $\alpha_{NT} \, \sim \,$ -0.8, and thermal bremsstrahlung, with thermal spectral index $\alpha_{T} \, \sim \,$ -0.1 \citep[\eg,][]{1992ARA&A..30..575C}; an illustration of this model is provided in Figure \ref{figToon1}. 
Prior surveys of spiral galaxies have reported  narrow distributions of spectral indices: \citet{1975AJ.....80..771S} found $\alpha$ = -0.85 between 1.4 and 5 GHz, \citet{1981A&A....94...29K} found $\alpha$ = -0.71 between 408 MHz and 10.7 GHz with a dispersion of 0.08 dex, and using a larger sample,  \citet{1982A&A...116..164G} found $\alpha$ = -0.74 with a dispersion of 0.12 dex over the same frequency range.

These sharply peaked empirical distributions of spectral indices are often interpreted as measures of the constancy of both the non-thermal spectral index and the thermal fraction  among sampled galaxies. 
These studies typically find that the spectra are well described by power laws over observed frequency ranges   spanning one or two orders of magnitude, a conclusion most evident from the ``average'' spectrum shown by \citet{1982A&A...116..164G}.  These results have led to the following common generalizations regarding the population of normal galaxies:  (1) the total radio spectral index is $\sim$~-0.7, (2) the synchrotron spectral index is  $\sim$~-0.8, and (3) synchrotron emission dominates over thermal emission at frequencies below 10 GHz, with the thermal fraction dropping to less than 10\% at frequencies $\scriptstyle \lesssim$ 1 GHz.  Furthermore, these generalizations are often extrapolated to model the radio spectrum at frequencies lower and higher than the data upon which they are based.

In the simple model with power-law components, \eg, Figure~\ref{figToon1}, the total spectrum flattens with increasing frequency as the thermal component becomes the dominant emission mechanism.
\citet{1997A&A...322...19N} use this curvature to fit for the thermal fraction at 1 GHz and find 8\% $\pm$ 1\% for a sample of 74 Shapley-Ames galaxies.  However, a similar study by \citet{1988AJ.....96...81D} of 41 spiral galaxies reports a much larger variation in thermal fraction.  This spectral decomposition method is  difficult,  since the power-law models often used may not be good approximations of the emission components.   One such case against a power-law model is presented by \citet{2010ApJ...710.1462W}, who favor a curved synchrotron spectrum to model the data.  Additionally, the synchrotron spectral index may be significantly steeper in some sources \citep{1997A&A...322...19N} than the mean value typical of  the galaxy population.

Other departures from the simple model are evident at low radio frequencies ($\nu\ \scriptstyle \lesssim$ 1 GHz), such as in the proto-typical starburst spectrum of M82  \citep{1992ARA&A..30..575C}.
In  studies involving larger samples of galaxies there have been  reports of flattening in the total spectrum \citep[\eg,][]{1972ApL....12...75S,1990ApJ...352...30I,1991A&A...250..302P}, but the physical interpretation remains uncertain.
Several plausible scenarios have been proposed including thermal absorption of a power-law synchrotron component or intrinsic curvature in the synchrotron spectrum.  For some ordinary galaxies, the case for thermal absorption appears to require unlikely physical parameters \citep{1991A&A...252..493P,1991A&A...251..442H} but plausible models have been presented for more extreme systems \citep[\eg,][]{2000ApJ...537..613A,2013MNRAS.431.3003L}.

Models of the synchrotron spectrum are often based on a population of relativistic electrons with a power-law distribution of energy $\gamma$: $q(\gamma) = q_{0} \gamma^{-s}$.  Several energy-dependent processes can alter the shape of this particle distribution and thereby alter the shape of the synchrotron spectrum.  
These processes include ionization and electronic excitation \citep{1975ApJ...196..689G}, inverse Compton, synchrotron radiation, and relativistic bremsstrahlung, although the latter is only weakly dependent on the particle's energy. 
The combined effect of these energy-dependent losses is that the initial power-law energy distribution (and resulting synchrotron spectrum) develops concave curvature, \ie, steeper at the higher energies (higher frequencies) and flatter at lower energies (lower frequencies).

A family of more elaborate models can be developed by considering the time evolution of the the synchrotron-emitting cosmic ray electron  (CRE) population, often by invoking a `single box'  with uniform particle  and energy density, or a `dynamical halo' model consisting of a 2-d galactic disk with perpendicular wind \citep[see][and references therein]{1990A&A...234..147P}.
For injection which is constant in time, these losses cause breaks of $\Delta s = 1.0$ which manifest as breaks of $\Delta \alpha =  0.5$ in the synchrotron spectrum---note that these breaks are not sharp but rather represent asymptotic values for which the transition occurs over at least an order of magnitude in frequency (for a graphical example, see \citealt{1991bja..book.....H}, p. 121; also, \citealt{1970PhRvD...2.2787J}).  
A low-frequency break (flattening with decreasing energy or frequency) is expected due to electronic excitation and ionization and a high-frequency break (steepening with increasing energy or frequency) results from inverse Compton and synchrotron radiative losses; under certain conditions, these breaks can fuse into a single observable with $\Delta \alpha = 1.0$.   Additionally, energy-dependent CRE re-acceleration \citep{1991A&A...252..493P} and energy-dependent diffusion combined with advective escape  can present observable spectral curvature.

However, quantitative conclusions drawn from these models remain questionable since they do not consider the high degree of inhomogeneity 
in real galaxies \citep[\eg,][]{1998MNRAS.300...30L, 2013A&A...557A.129T}.  Specifically, spatial variations in physical quantities such as density, magnetic field, cosmic ray injection, and interstellar radiation will result in spatial variations in the radio spectrum. Since the integrated spectrum is the sum over these different regions, the spectral features predicted by simple models will be smeared out over a larger frequency range and it becomes difficult to relate characteristics of the integrated spectrum to a single set of physical quantities. 
Due to these uncertainties, it is advantageous to explore additional relationships to better understand the nature of radio sources in galaxies.  Some trends have already been identified in previous studies, such as a relation between spectral index and galaxy morphological type 
 \citep[][wherein early types have slightly flatter spectra]{1991AJ....101..362C,1975AJ.....80..771S},  between spectral index and optical color \citep[][wherein redder galaxies are slightly flatter]{1993ApJ...410..626D}, and spectral index versus optical axial ratio \citep{1990ApJ...352...30I}, although the latter two results have not been well established.

It is the goal of this paper to better establish the shape of the integrated radio continuum spectrum below 5 GHz  and to test for relationships between source properties. 
In order to achieve a substantially larger sample of galaxies than previous efforts, and to better guard against selection effects and background source confusion, we have designed a new study to investigate the nature of  low-frequency radio spectra using a suite of  high-resolution survey data. 
We describe the properties of these surveys and  the  sample selection strategy in Section~\ref{secSample}.  Survey measurement techniques and derivation of additional source properties are detailed in Section~\ref{secData}.  The measured spectral indices and their relationships with  source properties and ancillary data are presented in Section~\ref{secResult} and discussed in Section~\ref{secDisc}.

\section{Survey Properties and Sample Selection}\label{secSample}

We selected four northern-hemisphere large-area radio surveys as the basis for the  spectral data: the VLA Low-Frequency Sky Survey Redux (VLSSr; \citet{2012RaSc...47.....L}), the Westerbork Northern Sky Survey
 (WENSS;  \citet{1997A&AS..124..259R}), the NRAO VLA Sky Survey (NVSS;  \citet{1998AJ....115.1693C}), and the Green Bank 6 cm Survey (GB6; \citet{1996ApJS..103..427G}).   Table~\ref{table-survey-prop} summarizes key properties of these surveys and Figure~\ref{figToon1} demonstrates how they sample the radio spectrum, from 74 MHz to 4.85 GHz, in roughly equal intervals.
The VLSSr, a re-reduction of the original VLSS survey data  \citep{2007AJ....134.1245C}  incorporating enhanced data processing and imaging algorithms, provides a critical improvement in sensitivity necessary to detect many of the sources in our sample.

\begin{deluxetable*}{lccccc}
\tabletypesize{\scriptsize}
\tablecaption{Survey Properties.\label{table-survey-prop}}
\tablewidth{0pt}
\tablehead{
\colhead{Survey} & \colhead{Frequency} & \colhead{Coverage} & \colhead{Resolution} & \colhead{LAS\tablenotemark{a}} & \colhead{Sensitivity} \\
\colhead{Name} & \colhead{(MHz)} & \colhead{($\degr$)} & \colhead{($''$)} & \colhead{($'$)} & \colhead{(mJy beam$^{-1}$)}
}
\startdata
VLSSr & 74& $\delta >$ -30 & 75 $\times$ 75 & 13.3 &54\tablenotemark{b} \\
WENSS & 325& $\delta >$ +28.5 & 54 $\times$ 54/$\sin\delta$ & 60 & 3.6 \\
NVSS & 1400& $\delta >$ -40 & 45 $\times$ 45 & 15 & 0.45 \\
GB6 & 4850 &  +75 $> \, \delta >$ 0  & 216 $\times$ 204 & 10 & 3.8\tablenotemark{c} \\
\enddata
\tablenotetext{a}{The largest angular scale to which the survey is sensitive.}
\tablenotetext{b}{The median source-free RMS measured from the survey cutouts used in this project.}
\tablenotetext{c}{The average over the declination-dependent sensitivity curve appearing in \citet{1996ApJS..103..427G}  for the declination range  $+35 \leq \delta \leq +75$. }
\end{deluxetable*}

We restrict this project to an annulus in declination, $+35 \leq \delta \leq +75$, which is covered by all surveys and for which the WENSS declination-dependent resolution element is not very distorted.  
All entries in the 3rd Catalog of Bright Galaxies (RC3; \citet{1991rc3..book.....D}) inside this declination range, with optical magnitude (B-band, 400-500 nm) B$_\mathrm{T} <14$ and  optical major axis $<10'$ are compiled, resulting in a sample size of 787. Note that this restriction on optical size is important to ensure accurate measurements of the total flux from the radio surveys. 

Image cutouts ($15' \times 15'$) are obtained at the location of each optical position for each of the VLSSr, WENSS, and NVSS surveys and all radio data are scaled to the flux-density standard of \citet{1977A&A....61...99B}. 
Each source is first investigated using the NVSS survey images, for which the point source sensitivity and resolution are superior  to the other surveys.   Since background confusion is known to be a serious concern (the NVSS catalog source density is 50 per square degree brighter than 2 mJy) we adopt a radio-optical coherence length of $30''$ to reject background sources with 99\% confidence \citep{1998AJ....115.1693C}. 
Additionally, we want to avoid background objects which could create confusion for the lower resolution surveys, so we reject sources having multiple radio components within $2'$ of the optical position.  A total of 197 galaxies are rejected on the basis of coherence length or multiple components.

We also  restrict our sample to include only radio sources for which the NVSS-WENSS 2-point spectral index can be well determined. Based on simulated measurements with  a standard $\alpha = -0.7$ spectral index, in order to achieve a measurement uncertainty $\Delta\alpha_{2pt} \le 0.1$ we need to consider WENSS sources $\ge 7\sigma$ (NVSS sources $\ge 20\sigma$).  Furthermore, by setting a demanding NVSS flux threshold, we can expect to detect these sources in the WENSS survey even if they have atypical spectra. Had we set a lower flux threshold, flat spectrum sources would have been too faint to be detected in WENSS, leading to a preference for steeper spectrum sources in the final sample.  
A total of 321 galaxies are rejected for not having radio counterparts bright enough for our spectral investigation. 
A small number of additional galaxies are rejected due to issues obtaining one or more of the survey cutouts, or because the  cutouts are severely contaminated by artifacts from a bright nearby source. After these selections, 250 galaxies remain in our final sample.

\section{Data}\label{secData}
\subsection{Survey Measurements}\label{secMeasure}

Since many galaxies in our sample have optical sizes larger than the survey resolution element, a 2-d Gaussian fit at the full survey resolution may not provide the best estimate of the total flux \citep{2008AJ....136.1889O}.  
Accordingly, the NVSS and WENSS  survey cutouts are convolved with Gaussian beams 2, 3, and 4 times larger than the native survey resolution to produce lower resolution images. The radio source is fit using the {\it AIPS} task {\tt JMFIT} for each scale; if the source size is resolved and the integrated flux is at least 10\% larger than the peak flux then the integrated flux is recorded, otherwise the peak flux is used. The final flux is taken  from the scale which produces the greatest signal-to-noise ratio while maintaining coincidence with the optical position.  This method produces values which are on average 5.4\% greater than those listed in the NVSS catalog and 4.5\% greater than values in the WENSS catalog. 

The VLSSr peak flux is measured from the full resolution survey cutout at the position of the optical source and corrected for beam resolution using the measured NVSS size.  The source-free RMS in the VLSSr cutout is used to determine the uncertainty in the VLSSr flux.  The GB6 fluxes and errors are compiled from the published catalog using entries within $1'$ of the optical positions. Sources having no GB6 association are assigned 5$\sigma$ upper limits using the GB6 declination-dependent sensitivity.  

Using these methods, we record NVSS fluxes for 250 sources, WENSS fluxes for 233 sources, VLSSr fluxes (greater than 3$\sigma$) for 89 sources, and GB6 fluxes (greater than 5$\sigma$) for 85 sources.  The upper limits are also recorded and used in further statistical analysis.  The survey measurements are listed in their entirety in Table \ref{table-data1}.  The nine columns of Table \ref{table-data1} give the following information:\\
{\bf \indent Column 1.} Uppsala General Catalog (UGC) number \citep{1973ugcg.book.....N}. \\
{\bf \indent Columns 2, 3.}  NVSS flux and uncertainty in mJy, measured from the survey cutout images using the \textit{AIPS} task {\tt JMFIT}, selected from multiple resolutions.\\
{\bf \indent Columns 4, 5.} NVSS fitted size in arcseconds and position angle in degrees, nominal full-width at half max, from {\tt JMFIT}.    \\
{\bf \indent Columns 6, 7} WENSS flux and uncertainty in mJy, measured from the survey cutout images using the \textit{AIPS} task {\tt JMFIT}, selected from multiple resolutions.\\
{\bf \indent Columns 8, 9.} VLSSr flux and error in mJy, measured from the survey cutout images.  Flux values are determined by taking the image value at the location of the optical source center and correcting for the ratio of peak to integrated flux based on the measured NVSS size.  Flux uncertainties are the source-free RMS of the cutout image multiplied by the same size correction factor that was used for the flux.  \\

\begin{deluxetable*}{ccccc@{\extracolsep{5pt}}c@{\extracolsep{-5pt}}c@{\extracolsep{5pt}}c@{\extracolsep{-5pt}}c@{\extracolsep{5pt}}c@{\extracolsep{-5pt}}c}
\tabletypesize{\scriptsize}
\tablecaption{Survey Measurements.\label{table-data1}}
\tablewidth{0pt}
\tablehead{
\colhead{}&\multicolumn{4}{c}{NVSS Quantities}&\multicolumn{2}{c}{WENSS Quantities}&\multicolumn{2}{c}{VLSSr Quantities}&\multicolumn{2}{c}{GB6 Quantities}\\
\cline{2-5} \cline{6-7}\cline{8-9}\cline{10-11}\\[-7pt]
\colhead{Galaxy Name}&\colhead{Flux}&\colhead{Error}&\colhead{$\theta_M \times \theta_m$}&
\colhead{P. A.} & \colhead{ Flux} & \colhead{ Error} &
\colhead{ Flux} & \colhead{ Error}&\colhead{ Flux} & \colhead{ Error} \\
\colhead{(UGC \#)} & \colhead{(mJy)} & \colhead{(mJy)} & \colhead{($''\,  \times\, ''$)}  &
\colhead{($\degr$)} & \colhead{(mJy)} & \colhead{(mJy)} &
\colhead{(mJy)} & \colhead{(mJy)}&\colhead{(mJy)} & \colhead{(mJy)} \\
\colhead{(1)} & \colhead{(2)} & \colhead{(3)} & \colhead{(4)} & \colhead{(5)} &
\colhead{(6)} & \colhead{(7)} & \colhead{(8)} &
\colhead{(9)} & \colhead{(10)} & \colhead{(11)} }
\startdata
444 & 10.0 & 1.1 & 37 $\times$ 26 & 81 & 24 & 2 & $<$161 & \nodata & $<$19 & \nodata \\ 
480 & 38.4 & 1.6 & 40 $\times$ 38 & 125 & 109 & 4 & 311 & 61 & $<$19 & \nodata \\ 
528 & 142 & 4 & 48 $\times$ 45 & 79 & 262 & 10 & 444 & 63 & 58 & 6 \\ 
625 & 19.6 & 1.2 & 45 $\times$ 31 & 134 & 42 & 5 & 287 & 75 & $<$18 & \nodata \\ 
758 & 20.6 & 0.9 & 26 $\times$ 17 & 72 & 52 & 4 & 189 & 50 & $<$19 & \nodata \\ 
1111 & 31.8 & 1.0 & 14 $\times$ 10 & 42 & 64 & 4 & $<$173 & \nodata & $<$19 & \nodata \\ 
1220 & 19.3 & 0.8 & 25 $\times$ 13 & 22 & 45 & 4 & $<$173 & \nodata & $<$19 & \nodata \\ 
1347 & 13.0 & 1.1 & 43 $\times$ 32 & 104 & 31 & 6 & $<$210 & \nodata & $<$19 & \nodata \\ 
1348 & 67.4 & 2.4 & 44 $\times$ 11 & 85 & 263 & 10 & 840 & 62 & 32 & 4 \\ 
1355 & 18.5 & 0.9 & 23 $\times$ 21 & 39 & 37 & 4 & $<$244 & \nodata & $<$18 & \nodata \\ 
\enddata
\tablecomments{Table \ref{table-data1} is published in its entirety in the electronic edition of the {\it Astronomical Journal}.  A portion is shown here for guidance regarding its form and content.}
\end{deluxetable*}

\subsection{Additional Source Properties}\label{addProp}

We compile a suite of additional data for each source based on available optical, infrared and radio information, which will be tested for relationships with radio spectral properties in Sections \ref{secHubData} and \ref{secAddData}.  
The optical morphology is assessed using the numerical Hubble stage T, a  progression from early to late types, as recorded in the RC3 catalog.  
Values for the position, optical major axis (D25), optical axial ratio,
B-band magnitude, B-V and U-B colors, 
and systemic velocity are also taken from the RC3 catalog.

Infrared fluxes at 60 and 100 $\micron$ are extracted from the following IRAS databases, in order of preference: \textit{The Bright Galaxy Sample} \citep{1989AJ.....98..766S}, \textit{Large Optical Galaxies} \citep{1988ApJS...68...91R}, \textit{Small-Scale Structure Catalog} \citep{1988SSSC..C......0H}, \textit{Faint Source Catalog} \citep{1992ifss.book.....M}, and the \textit{Point Source Catalog} \citep{1988iras....1.....B}.   Only IRAS sources within $2'$ of the optical positions are considered. 
The 60 and 100 $\micron$ fluxes are combined into a single quantity, $FIR$, which estimates the flux between 42.5 and 122.5 $\micron$ \citep{1985ApJ...298L...7H},
\begin{equation}\label{eqn-FIR}
FIR = 1.26 \times 10^{-14}\, \left(2.58 \,S_{60\, \micron} + S_{100\, \micron}\right)  \,\,\, \mathrm{W \,m^{-2}}
\end{equation}
where $S_{60\, \micron}$ and $S_{100\, \micron}$ are the 60 and 100 $\micron$ fluxes in Jy.

We adopt the relation between star formation rate (SFR) and radio free-free luminosity, $L_{\mathrm{th}}$, given in \citet{2011ApJ...737...67M}
\begin{eqnarray}\label{murSFR}
\frac{SFR}{\mathrm{M}_{\sun}\, \mathrm{yr}^{-1}} = 4.6\,\times&&10^{-28} \bigg( \frac{L_{\mathrm{th}}}{\mathrm{ergs\,\, s^{-1} Hz^{-1}}} \bigg) \nonumber  \\ \times &&
\bigg( \frac{T_{e}}{10^4\, \mathrm{K}} \bigg)^{-0.45} \, \bigg( \frac{\nu}{\mathrm{GHz}} \bigg)^{0.1} 
\end{eqnarray}
and the relation between SFR and IR luminosity, $L_{\mathrm{IR}}$, from \citet{1998ApJ...498..541K}
\begin{equation}\label{kenSFR}
\frac{SFR}{\mathrm{M}_{\sun}\, \mathrm{yr}^{-1}} = 4.5\times10^{-44} \bigg( \frac{L_{\mathrm{IR}}}{\mathrm{ergs\,\, s^{-1}}}\bigg)
\end{equation}
By equating Equation~\ref{murSFR} and Equation~\ref{kenSFR} we derive an expression for the optically-thin thermal \textit{radio} flux, $^{\mathrm{IR}}S_{\mathrm{th}}$, 
\begin{eqnarray}\label{eqn-1}
^{\mathrm{IR}}S_{\mathrm{th}}(\nu) = 1.4\,\times&&10^{10}\,  \bigg( \frac{FIR}{\mathrm{W \,m^{-2}}} \bigg)\, \nonumber  \\  \times &&
\bigg( \frac{T_{e}}{10^4\, \mathrm{K}} \bigg)^{0.45} \, \bigg( \frac{\nu}{\mathrm{GHz}} \bigg)^{-0.1}\,\,\,  \mathrm{Jy}
\end{eqnarray}
for which we have assumed $L_{\mathrm{IR}} (8-1000 \,\mu \mathrm{m}) = 1.5 L_{\mathrm{FIR}} (42.5-122.5 \,\mu \mathrm{m})$ as in \citet{2001ApJ...554..803Y}.
We introduce the superscript `IR' to remind the reader that this quantity is derived from the far-IR flux using simple assumptions. 
 We adopt an electron temperature of 10$^4$ K and tabulate the thermal radio flux at 1.4 GHz, $^{\mathrm{IR}}S_{\mathrm{th, 1.4}} = {^{\mathrm{IR}}S_{\mathrm{th}}} (1.4~\mathrm{GHz})$. Note that because Equations \ref{murSFR} and \ref{kenSFR} are based on starburst population synthesis models with somewhat different assumptions, the numerical coefficient of Equation \ref{eqn-1} has some additional uncertainty.

We define the thermal fraction, \ftherm, as the ratio of thermal radio flux to total radio flux, 
\begin{equation}\label{eqn-ftherm}
f_{\mathrm{th}}(\nu) = S_{\mathrm{th}}(\nu) / S(\nu)
\end{equation}
We tabulate the thermal fraction at 1.4 GHz, $^{\mathrm{IR}}$\fthermL, using  $^{\mathrm{IR}}S_{\mathrm{th, 1.4}}$ as the thermal radio flux and our measurement of the NVSS flux as the total radio flux at 1.4 GHz,
\begin{equation}\label{eqn-ftherm14}
{^{\mathrm{IR}}f_{\mathrm{th,1.4}}} = {^{\mathrm{IR}}S_{\mathrm{th,1.4}}} / S_\mathrm{NVSS}
\end{equation}

The radio-FIR ratio $q$ is calculated from the IRAS data and our NVSS survey measurements following \citet{1985ApJ...298L...7H},
\begin{equation}\label{eqn-q}
q = \log \left( \frac{FIR}{3.75 \times 10^{12} \,\,\mathrm{W \,\, m^{-2}}} \right) -  \log \left( \frac{S_{1.4 \,\mathrm{GHz}}}{\mathrm{W \,\, m^{-2} \,\, Hz^{-1}}} \right)
\end{equation}
Since the expressions for $^{\mathrm{IR}}$\fthermL\ and $q$ both contain the ratio of $FIR$ to the total radio flux, $^{\mathrm{IR}}$\fthermL\ can be expressed as a function of $q$, 
\begin{equation}\label{eqn-F-q}
{^{\mathrm{IR}}f_{\mathrm{th,1.4}}} = 1.7 \times 10^{\,(q -  3.53)}
\end{equation}
Based on this relation sources with small values of $q$, which would classically be described as having a radio excess, could alternatively be interpreted as having a low thermal fraction.  Note that values of $q$ more extreme than 3.29 would be difficult to interpret in this manner, since their thermal fractions would exceed unity.  However, not only does our study not contain any sources with $q$ this extreme, but neither does the study of \citet{2001ApJ...554..803Y} which includes more than 1800 sources.

\begin{deluxetable*}{cccccccccc}
\tabletypesize{\scriptsize}
\tablecaption{Far-IR Data and Derived Quantities.\label{table-data3}}
\tablewidth{0pt}
\tablehead{
\colhead{Galaxy Name} & \colhead{$S_{60 \micron}$} & \colhead{$S_{100 \micron}$} & \colhead{$FIR$ }&\colhead{$^{\mathrm{IR}}S_{\mathrm{th,1.4}}$}&\colhead{$^{\mathrm{IR}}$\fthermL}  &\colhead{R-FIR $q$}\\
\colhead{(UGC \#)} & \colhead{(Jy)} & \colhead{(Jy)} & \colhead{($10^{-14}$ W m$^{-2}$)}  & \colhead{(mJy)} & \colhead{(\%)} & \colhead{} \\
\colhead{(1)} & \colhead{(2)} & \colhead{(3)} & \colhead{(4)} & \colhead{(5)} &
\colhead{(6)} & \colhead{(7)} } 
\startdata
444 & 0.8 & 2.2 & 5.4 & 0.7 & 7 & 2.16 \\ 
480 & 1.4 & 3.6 & 9.0 & 1.1 & 3 & 1.79 \\ 
528 & 25 & 44 & 137 & 17 & 12 & 2.41 \\ 
625 & 2.4 & 6.6 & 16 & 2.0 & 10 & 2.34 \\ 
758 & 2.0 & 5.2 & 13 & 1.7 & 8 & 2.24 \\ 
1111 & 2.0 & 3.5 & 11 & 1.4 & 4 & 1.96 \\ 
1220 & 1.9 & 4.6 & 12 & 1.5 & 8 & 2.22 \\ 
1347 & 1.5 & 3.8 & 9.7 & 1.2 & 10 & 2.30 \\ 
1348 & 0.2 & 0.9 & 1.9 & 0.2 & $<$1 & 0.88 \\ 
1355 & 3.4 & 6.2 & 19 & 2.4 & 13 & 2.43 \\ 
\enddata
\tablecomments{Table \ref{table-data3} is published in its entirety in the electronic edition of the {\it Astronomical Journal}.  A portion is shown here for guidance regarding its form and content.}
\end{deluxetable*}

The infrared fluxes and derived quantities are listed in  Table \ref{table-data3}.  The seven columns of Table \ref{table-data3} give the following information:\\
{\bf \indent Column 1.} UGC number. \\
{\bf \indent Columns 2, 3.} The 60 and 100 $\micron$ IRAS fluxes used  in this study. \\  
{\bf \indent Column 4.} The total Far-IR emission between 42.5 and 122.5 $\micron$, calculated with Equation~\ref{eqn-FIR}. \\
{\bf \indent Column 5.}  The thermal flux in mJy calculated with Equation~\ref{eqn-1}.  \\
{\bf \indent Column 6.}  The thermal fraction at 1.4 GHz, calculated with Equation~\ref{eqn-ftherm14}. \\ 
{\bf \indent Column 7.}  The radio-FIR ratio $q$, calculated with Equation~\ref{eqn-q}. \\

Additional quantities are tabulated for the radio morphology and luminosity. 
We characterize the nature of the radio source by its compactness, defined as the ratio of the fitted NVSS major axis to the optical major axis, and by the radio surface brightness, using the ratio of the NVSS flux to the NVSS solid angle defined by the fitted source size.  
The Hubble flow distance is calculated from the RC3 velocity (galactic standard of rest) by adopting a value of $H_{\circ} = 70$ km s$^{-1}$ Mpc$^{-1}$.  Radio and infrared luminosities are estimated in the standard fashion, $L = 4\pi D^2S$,  for distance $D$ and flux $S$.  Hubble flow distances less than 20 Mpc and any derived luminosities are not considered in further analysis due to their large uncertainty.

A selection of source properties are listed in  Table \ref{table-data4}.  The eight columns of Table \ref{table-data4} give the following information:\\
{\bf \indent Column 1.} UGC number. \\
{\bf \indent Column 2.} Alternate source name. \\
{\bf \indent Column 3.} Optical position from RC3 \\ 
{\bf \indent Column 4.} Optical axial ratio (minor to major) derived from RC3.  \\
{\bf \indent Column 5.} Numerical Hubble type from RC3, from early to late morphologies.  Ellipticals are in the range -6 to -4, lenticulars -3 to -1,  spirals (from Sa to Sd)  0 to 8, Magellanic spirals and irregulars 9 and 10, respectively, compact irregulars, 11, non-Magellanic irregulars 90, and peculiars 99.  \\
{\bf \indent Column 6.} Log of the surface brightness, calculated as the NVSS flux divided by the NVSS solid angle (major axis diameter by minor axis diameter), in units of mJy per square arcminute.\\
{\bf \indent Column 7.} Radio compactness parameter, the ratio of NVSS major axis (Gaussian full-width at half max) to RC3 optical major axis (D25). \\
{\bf \indent Column 8.} The Hubble flow distance in Mpc. \\

\begin{deluxetable*}{cccccccc}
\tabletypesize{\scriptsize}
\tablecaption{Additional Source Properties.\label{table-data4}}
\tablewidth{0pt}
\tablehead{
\colhead{Galaxy Name} & \colhead{Alt. Name } & \colhead{Position (J2000)} & \colhead{Axial Ratio } &
\colhead{Type }&\colhead{$\log \Sigma$} &\colhead{compact} & \colhead{D (Mpc)} \\
\colhead{(1)} & \colhead{(2)} & \colhead{(3)} & \colhead{(4)} & \colhead{(5)} &
\colhead{(6)} & \colhead{(7)} & \colhead{(8)}  
}
\startdata
444 & \nodata & 00 42 04.7 \, 36 48 15 & 0.69 & \nodata & 1.6 & 0.54 & 155 \\ 
480 & NGC 218 & 00 46 32.0 \, 36 19 33 & 0.76 & \nodata & 2.0 & 0.43 & 162 \\ 
528 & NGC 278 & 00 52 04.6 \, 47 33 00 & 0.95 & 3 & 2.4 & 0.38 & \nodata \\ 
625 & IC 65 & 01 00 55.8 \, 47 40 51 & 0.30 & 4 & 1.7 & 0.19 & 40 \\ 
758 & NGC 425 & 01 13 02.5 \, 38 46 07 & 0.93 & \nodata & 2.2 & 0.41 & 93 \\ 
1111 & NGC 591 & 01 33 31.4 \, 35 40 07 & 0.83 & 0 & 2.9 & 0.18 & 67 \\ 
1220 & NGC 662 & 01 44 35.4 \, 37 41 48 & 0.63 & \nodata & 2.3 & 0.50 & 83 \\ 
1347 & \nodata & 01 52 45.8 \, 36 37 07 & 0.87 & 5 & 1.5 & 0.57 & 81 \\ 
1348 & NGC 708 & 01 52 46.4 \, 36 09 06 & 0.83 & -5 & 2.7 & 0.24 & 71 \\ 
1355 & \nodata & 01 53 36.5 \, 43 57 58 & 0.87 & 3 & 2.1 & 0.28 & 91 \\ 
\enddata
\tablecomments{Table \ref{table-data4} is published in its entirety in the electronic edition of the {\it Astronomical Journal}.  A portion is shown here for guidance regarding its form and content.}
\end{deluxetable*}

\section{Results}\label{secResult}

\begin{deluxetable*}{ccccccccccc}
\tabletypesize{\scriptsize}
\tablecaption{Spectral Properties.\label{table-data2}}
\tablewidth{0pt}
\tablehead{
\colhead{UGC \#} & \colhead{\alow } & \colhead{$\Delta$\alow } & \colhead{\amid } & \colhead{$\Delta$\amid }& \colhead{\ahigh } & \colhead{$\Delta$\ahigh }&\colhead{\flow }& \colhead{$\Delta$\flow }  \\
\colhead{(1)} & \colhead{(2)} & \colhead{(3)} & \colhead{(4)} & \colhead{(5)} &
\colhead{(6)} & \colhead{(7)} & \colhead{(8)} & \colhead{(9)} 
}
\startdata
444 & $>$-1.28 & \nodata & -0.60 & 0.09 & $<$0.50 & \nodata & -0.3 & 0.1 \\ 
480 & -0.71 & 0.14 & -0.71 & 0.04 & $<$-0.58 & \nodata & 2.9 & 0.6 \\ 
528 & -0.36 & 0.10 & -0.42 & 0.03 & -0.72 & 0.09 & 1.7 & 0.2 \\ 
625 & -1.29 & 0.20 & -0.53 & 0.10 & $<$-0.06 & \nodata & 6.8 & 2.0 \\ 
758 & -0.87 & 0.19 & -0.63 & 0.06 & $<$-0.08 & \nodata & 3.6 & 1.0 \\ 
1111 & $>$-0.67 & \nodata & -0.48 & 0.04 & $<$-0.43 & \nodata & 2.3 & 0.9 \\ 
1220 & $>$-0.92 & \nodata & -0.57 & 0.06 & $<$-0.03 & \nodata & 1.0 & 1.3 \\ 
1347 & $>$-1.29 & \nodata & -0.60 & 0.15 & $<$0.30 & \nodata & 3.5 & 2.4 \\ 
1348 & -0.78 & 0.06 & -0.93 & 0.04 & -0.60 & 0.10 & 3.2 & 0.3 \\ 
1355 & $>$-1.27 & \nodata & -0.47 & 0.08 & $<$-0.01 & \nodata & 0.2 & 2.2 \\ 
\enddata
\tablecomments{Table \ref{table-data2} is published in its entirety in the electronic edition of the {\it Astronomical Journal}.  A portion is shown here for guidance regarding its form and content.}
\end{deluxetable*}

\subsection{Spectral Properties of the Sample} 

The goal of this section is to present our spectral measurements and to describe the average spectrum of the entire sample.  
We calculate the spectral index between adjacent frequency pairs, when sources are detected in both surveys, or constrain the spectral index in the case that the source is only detected in one survey in the pair.  Spectral index limits are calculated  between the detection and the value of the flux upper limit; no value is produced when the source is not detected in either survey in the pair. Due to dissimilar survey sensitivities, many sources detected in the WENSS survey are not detected in the VLSSr, leading to lower limits on the low-frequency spectral index, and many sources detected in the NVSS survey are not detected in GB6, leading to upper limits on the high-frequency spectral index. 

We label these spectral indices \alow, \amid\ and \ahigh\ to represent their relationship to our frequency sampling as illustrated in Figure~\ref{figToon1}, and provide
these spectral measurements in  Table \ref{table-data2}.  The nine columns of Table \ref{table-data2} give the following information:\\
{\bf \indent Column 1.} UGC number. \\
{\bf \indent Columns 2, 3.} Spectral index \alow\ and uncertainty measured between the VLSSr and WENSS data. Limits are given when only one flux is detected, calculated as the slope between the detection and the flux upper limit.  No value is given when there are non-detections at both bands. \\
{\bf \indent Column 4, 5.}  Spectral index \amid\ and uncertainty measured between the WENSS and NVSS data.  Non-detections are treated as in columns 2 and 3. \\  
{\bf \indent Column 6, 7.}  Spectral index \ahigh\ and uncertainty measured between the NVSS and GB6 data. Non-detections are treated as in columns 2 and 3.\\
{\bf \indent Column 8, 9.}  The ratio and uncertainty between the VLSSr and WENSS flux measurements.  Negative values are due to the VLSSr flux measurement technique.  No value is given when the WENSS flux is not detected.  \\

After calculating the spectral index, our next goal is to characterize the distributions of \alow, \amid\ and \ahigh. When dealing with flux upper limits, we employ the \textit{Nondetects and Data Analysis} (\texttt{NADA})
package from the [R] computer language.
Specifically, the Akritas-Theil-Sen (ATS) nonparametric line is used to determine the population mean, which considers both detections and censored data using survival analysis techniques; the uncertainty in the population mean is estimated by nonparametric bootstrap.

\begin{figure}
\epsscale{1.20}
\plotone{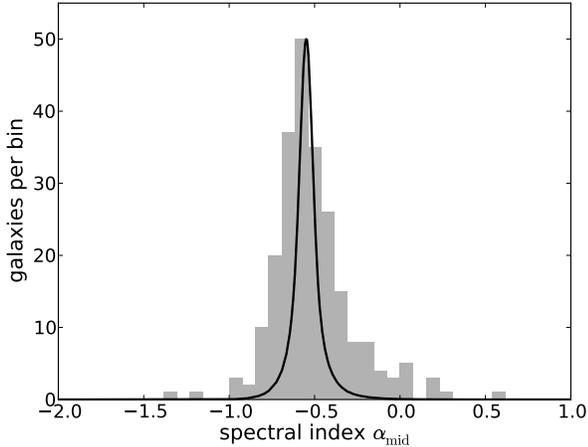}
\caption{The distribution of spectral index \amid, calculated between the WENSS and NVSS survey data, is shown for the sample of 250 sources (shaded).  These data have mean value  -0.55 and  standard error  0.01,  calculated after outlier rejection. The uncertainties in the flux measurements are used to simulate the  spectral index distribution which we would expect to observe if every source's intrinsic spectral index were equal to that of the sample mean.
The dispersion in  \amid\ is  substantially broader than can be explained by measurement uncertainty alone (solid line), indicating that much of the observed variability is intrinsic to the source population.} 
\label{figAmid}
\end{figure}

The most precise measure of the spectral index is between the 325 and 1400 MHz data (WENNS - NVSS, referred to herein as \amid).  A value of \amid\ is available for each of the 250 sources and has a typical uncertainty of $\pm$0.1 (see Figure \ref{figAmid}).  We find a mean \amid\ of -0.55 and dispersion between 0.10 and 0.20 depending on the degree of outlier rejection, resulting in a standard error in the mean of approximately 0.01.
A simulation incorporating the measured flux uncertainties is used to estimate the effect of measurement error on the dispersion in \amid. 
By scaling each source to have the same spectral index as the mean of \amid\ and resampling \amid\ using a parametric uncertainty model, we construct the distribution  we would expect to observe if each source had the same spectral index (see the solid curve in Figure~\ref{figAmid}). 
Since the dispersion in this simulated distribution is approximately 0.05, we find that much of the observed dispersion in \amid\ is due to intrinsic variability within the population and can not be attributed to measurement error.

\begin{figure}
\epsscale{1.20}
\plotone{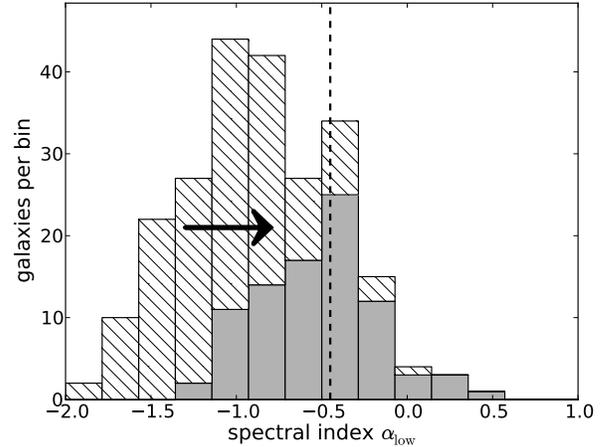}
\caption{The distribution of spectral index \alow, taken between the VLSSr and WENSS survey data, is shown for the sample of 250 sources.  Measurements (shaded) are computed from detections at both frequencies;  3$\sigma$ lower limits (hashed)  are derived from VLSSr non-detections.  The sample mean (dashed line) is determined to be $-0.45 \pm 0.05$,  incorporating both the detections and limits using survival analysis. The arrow overlaid on the hashed bins emphasizes the direction of the spectral index limits and therefore the influence these bins have on the sample mean.}
\label{figAlow}
\end{figure}

The spectral index between the lowest frequency data in our study, \alow, is calculated using flux measurements at 74 and 325 MHz (VLSSr - WENSS). We choose to censor the VLSSr flux measurements below 3$\sigma$ for the purposes of calculating the spectral index and instead use the 3$\sigma$ value to determine a lower limit to \alow.  We find a mean \alow\ for the sample of $-0.45 \pm 0.05$, incorporating both the detections and limits with survival analysis (see Figure \ref{figAlow}).  The dispersion, considering only the detections, is 0.37.   The low frequency flux ratio \flow, defined as the ratio of VLSSr to WENSS flux, is also calculated for each source. The advantage of \flow\ is that it has been tabulated along with an uncertainty for each source instead of being censored in the manner of \alow.

\begin{figure}
\epsscale{1.20}
\plotone{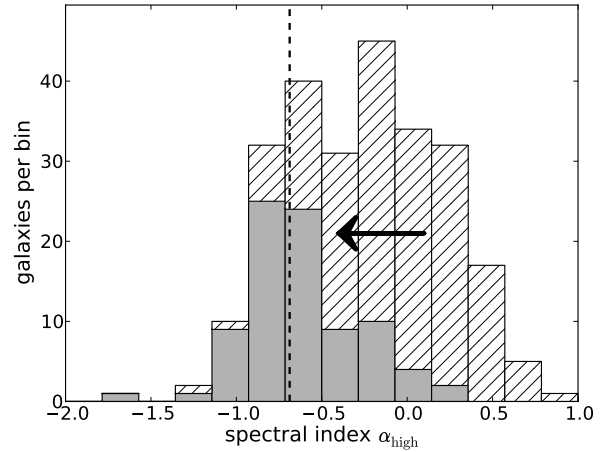}
\caption{The distribution of spectral index \ahigh, between the NVSS and GB6 survey data, for the sample of 250 sources.  Measurements  (shaded) are computed from detections at both frequencies;  5$\sigma$ upper limits (hashed) are based on NVSS detections for which no corresponding GB6 catalog association was identified.  The sample mean, $-0.69 \pm 0.04$,  is fit with survival analysis techniques to incorporate the detections and limits. The arrow overlaid on the hashed bins emphasizes the direction of the spectral index limits and therefore the influence these bins have on the sample mean.}
\label{figAhigh}
\end{figure}

\begin{figure}
\epsscale{1.20}
\plotone{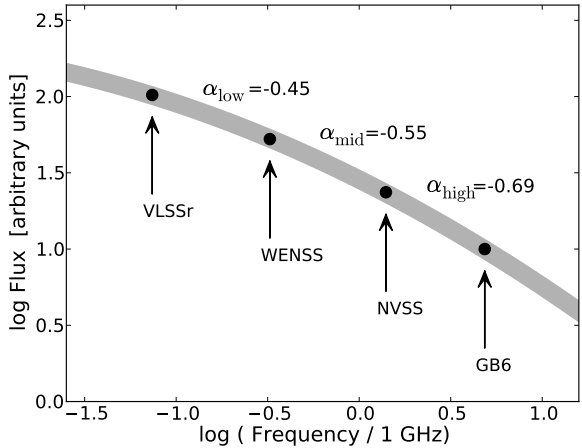}
\caption{Illustration summarizing the results of our spectral measurements for the galaxy sample.  The scaled mean flux for each radio survey (filled circles) is fit with a function having spectral index $\alpha$ = -0.7 at 10 GHz and curvature $\Delta \alpha$ = -0.2 per logarithmic frequency decade (shaded).   }
\label{figToon2}
\end{figure}

Between 1400 and 4850 MHz (NVSS - GB6) we calculate \ahigh, the spectral index between the highest frequency  pair.  The GB6 5$\sigma$ declination-dependent upper limits are used to produce corresponding upper limits in \ahigh\ when catalog fluxes are unavailable.  A sample mean of $-0.69 \pm 0.04$ (see Figure \ref{figAhigh}) is determined using survival analysis, and the dispersion of  the detections is 0.30.

The differences between mean values of \alow, \amid, and \ahigh\ are tested for significance against the t-distribution, where the variance and effective degrees of freedom are estimated following procedures for samples with unequal variances. 
The change from \alow\ to \amid\ is suggestive but not highly significant, at a level of 95\%, whereas the differences between \amid\ and \ahigh\ and between \alow\ and \ahigh\ are both deemed significant at a level greater than 99.9\%.  
Taken as a whole these results demonstrate that, across the entire frequency range of this study, the average spectrum of galaxies in our sample can not be well described by a single power law; this amounts to a detection of curvature among the general population in our sample.  The average spectrum can be approximated by a generic function with logarithmic curvature $\beta = -0.1$, where  $S(\nu) \propto \nu^{\alpha + \beta \log \nu}$.  For this function, the spectral index changes by a value $\Delta \alpha = 2 \beta$ for each logarithmic decade in frequency.
A summary of these results is illustrated in Figure \ref{figToon2}, which shows a spectrum derived from the mean spectral indices.

\begin{figure}
\epsscale{1.20}
\plotone{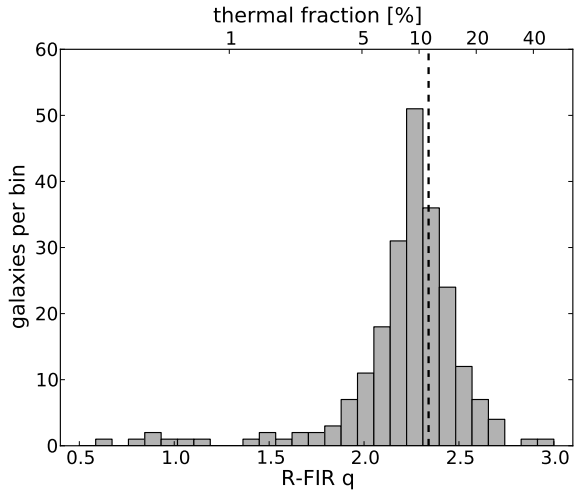}
\caption{The distribution of the radio to far-IR ratio $q$, calculated with Equation~\ref{eqn-q},  for a sample of 221 galaxies for which IRAS data was available. The mean value from \citet{2001ApJ...554..803Y},  $q=2.34$, is indicated by the dashed line.  Also labeled is the thermal fraction at 1.4 GHz, ${^{\mathrm{IR}}f_{\mathrm{th,1.4}}}$, calculated with Equation~\ref{eqn-ftherm14}.  The median value of $q$ for our sample is 2.27, which corresponds to a thermal fraction of 9\%.  
We attribute the displacement of these values of $q$ from those in  \citet{2001ApJ...554..803Y} to differences in sample selection: our sample is limited by radio flux and optical magnitude, and theirs is 60 $\micron$ flux limited.}
\label{figFtherm}
\end{figure}

\begin{figure}
\epsscale{1.20}
\plotone{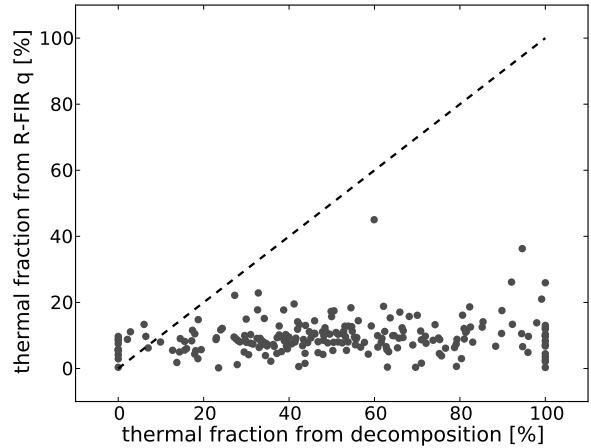}
\caption{This scatter plot is used to test for a correlation between the thermal fraction determined from two independent methods.
\textit{Ordinate}: the thermal fraction calculated with the IRAS and NVSS fluxes,  ${^{\mathrm{IR}}f_{\mathrm{th,1.4}}}$, as in  Figure \ref{figFtherm}. 
\textit{Abscissa}: the thermal fraction calculated from spectral decomposition,  ${^{\mathrm{R}}f_{\mathrm{th,1.4}}}$, from Equation~\ref{eqn-simp-model}. 
The dashed line indicates a  1:1 correlation, which would be expected if both methods produced accurate estimates of the thermal fraction; however, no such relationship is observed. }
\label{figFtherm-vs-q}
\end{figure}

\subsection{Calculations of the Thermal Fraction}\label{secCalcTherm}

In this section, two independent methods are used to calculate the thermal fraction, \fthermL, defined as the ratio of the free-free emission to the total radio continuum at 1.4 GHz, and the results are compared. The first method,  discussed in Section~\ref{addProp}, uses the IRAS and NVSS fluxes  to produce the values of $^{\mathrm{IR}}$\fthermL\ which appear in Table~\ref{table-data3}. The value of $^{\mathrm{IR}}$\fthermL\ can also be expressed as a function of the radio far-IR ratio $q$ as specified by Equation~\ref{eqn-F-q}. The distribution of these values of $q$, and the corresponding values of $^{\mathrm{IR}}$\fthermL\, are displayed in Figure~\ref{figFtherm}. 

The second method of calculating the thermal fraction is based on the model of power-law emission components, \ie, synchrotron and free-free having spectral indices $-0.8$ and $-0.1$, respectively, as illustrated with Figure~\ref{figToon1}. The functional form of this model can be written as in Equation~\ref{eqn-simp-model},
\begin{equation}\label{eqn-simp-model}
S(\nu) \propto f_{\mathrm{th}} \left( \frac{\nu}{\nu_0} \right) ^{-0.1} + \, (1- f_{\mathrm{th}}) \left( \frac{\nu}{\nu_0} \right) ^{-0.8}
\end{equation}
where $f_{\mathrm{th}}$ is the thermal fraction at a chosen reference frequency $\nu_0$. 
By evaluating this model at $\nu_1 = 325$ MHz and $\nu_2 = 1.4$ GHz we derive the following expression for  $^{\mathrm{R}}$\fthermL\ as a function of \amid:
\begin{equation}\label{eqn-specific-model}
{^{\mathrm{R}}f_{\mathrm{th}, 1.4}} = \frac{(\nu_1 / \nu_2)^{\alpha_{\mathrm{mid}}} - (\nu_1 / \nu_2)^{-0.8}}{(\nu_1 / \nu_2)^{-0.1} - (\nu_1 / \nu_2)^{-0.8}}  
\end{equation}
where the superscript `R' denotes that this value is derived solely from radio observations plus simple assumptions.
Equation~\ref{eqn-specific-model} represents a process which is often referred to as spectral decomposition.   Some applications of spectral decomposition, having available a larger number of frequency measurements across a wider bandwidth, have allowed the synchrotron spectral index to be a free parameter; here, we avoid this additional complexity in order to prevent a degenerate parameter space. Note that our model does not satisfactorily explain observed spectral indices flatter than -0.1 or steeper than -0.8, and as such we assign such observations thermal fractions of 100\% and 0\%, respectively.

In Figure~\ref{figFtherm-vs-q}, the distribution of $^{\mathrm{IR}}$\fthermL\ (from Equation~\ref{eqn-ftherm14}) is compared to the distribution of $^{\mathrm{R}}$\fthermL\  (calculated with Equation~\ref{eqn-specific-model}). 
We find no correlation between the thermal fractions estimated from these independent methods.  
Moreover, the values of $^{\mathrm{IR}}$\fthermL\ are narrowly distributed with a median of 9\% and dispersion 3\%, and agree well with the expected values from prior studies, while the values of $^{\mathrm{R}}$\fthermL\ are, on average,  much larger and have much greater variation (mean 51\% and dispersion 26\%). The values produced by spectral decomposition are unsatisfying for several reasons, most notably  the large mean and dispersion, and failure to properly account for  spectral indices outside the interval $-0.8$, $-0.1$.  This suggests that the spectral decomposition model given by Equation~\ref{eqn-simp-model} is an inadequate description of the spectra below 1 GHz.

\subsection{Relationships with Hubble Type}\label{secHubData}

We investigate relations between the tabulated source properties (\ie, spectral measurements, FIR-derived quantities and additional data) and the numerical Hubble type.  The data are binned by morphological types E, L, Sa, Sb, Sc, and Sd/I and the bin means and standard errors are used to determine the significance level of correlation. After reviewing a large number of possible correlations, we present in this section the results deemed most relevant to this study. 
 
\begin{figure}
\epsscale{1.20}
\plotone{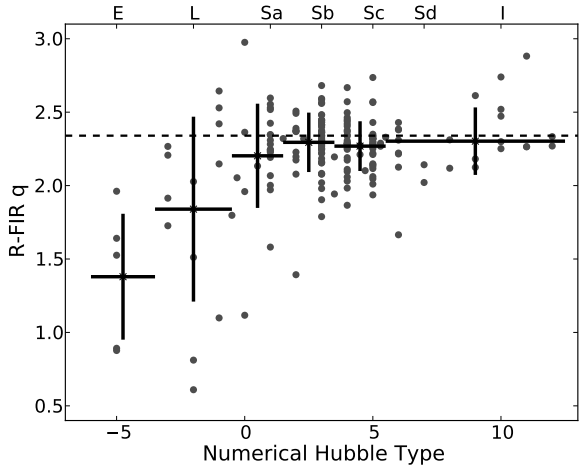}
\caption{A plot of the radio-farIR ratio $q$, as defined in Equation~\ref{eqn-q}, versus  the numerical  Hubble type. The crosses indicate the mean and standard deviation for bins containing E, L, Sa, Sb, Sc, and Sd/I types.  
The dashed line indicates the value $q=2.34$, a value often associated with star formation activity \citep{2001ApJ...554..803Y}; galaxies with $q$ values which are less than 1.64 belong to a radio-excess population commonly associated with AGN. The early-type galaxies in our sample have, on average, lower values of $q$ than later morphological types.}
\label{type-corr1}
\end{figure}

First, we discuss the observed relation between $q$ and numerical Hubble type, shown in Figure~\ref{type-corr1}.  Taken as a whole, the early-type (E and L) galaxies in our sample have significantly smaller values of $q$ than spiral types Sa through Sd. 
While the late types in our sample typically have values of $q$ consistent with the expected value for ordinary star-forming galaxies, many early types in our sample have  `radio-excess' values of $q$ which can be an indicator of AGN activity \citep{2001ApJ...554..803Y}.

\begin{figure}
\epsscale{1.20}
\plotone{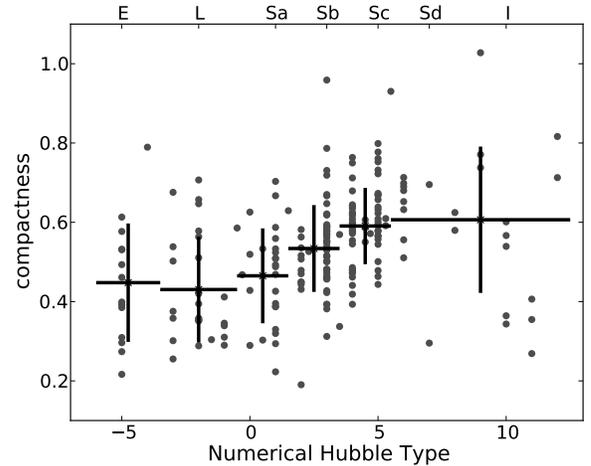}
\caption{The radio-optical compactness parameter, defined as the ratio of the radio major axis to the optical major axis, versus the numerical  Hubble type.   The crosses indicate the mean and standard deviation for bins containing E, L, Sa, Sb, Sc, and Sd/I types.  }
\label{type-corr2}
\end{figure}

\begin{figure}
\epsscale{1.20}
\plotone{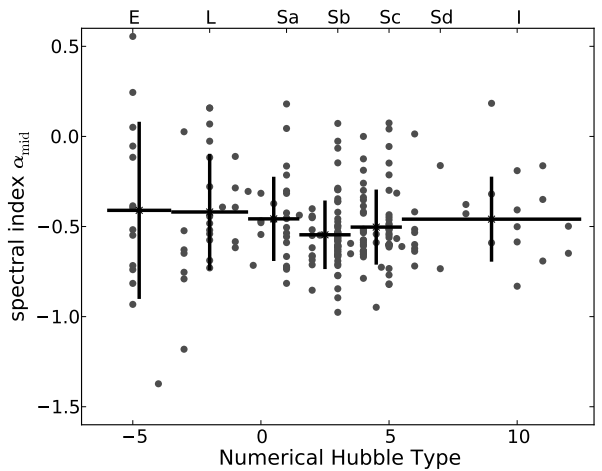}
\caption{This scatter plot is used to explore the relationship between spectral index \amid\ and morphological type, using the numerical Hubble type.  The crosses indicate the mean and standard deviation for bins containing E, L, Sa, Sb, Sc, and Sd/I types.  No appreciable differences are identified in the spectral index of different morphological types.}
\label{type-corr3}
\end{figure}

We  identify a relationship between the numerical Hubble type and our radio compactness parameter, the ratio of the radio to optical major axis diameter (see Figure~\ref{type-corr2}).  On average, the early-type galaxies in this sample harbor radio sources which are smaller with respect to the optical host as compared with later morphological types.  
There also exists substantial variation in compactness within the sub-population of spiral types Sa to Sd, where the later tends to be more extended than the former. 
One possible explanation is that the star formation in late types is more distributed throughout the optical disk but is more nucleated in early types;  alternatively, the radio emission from early types may be due in part to a low-luminosity AGN. 
A  relationship between compactness and radio surface brightness is also identified, although this may be due in part to selection effects from using a flux threshold.  

We test for a relationship between \amid\ and Hubble type, but do not identify  a significant  trend in our data (see Figure~\ref{type-corr3}). This is an unexpected result considering that some prior studies have detected such a relation \citep[\eg,][]{ 1991AJ....101..362C, 1993ApJ...410..626D}, although they typically used data at higher frequencies to calculate the spectral index.  Additionally, the distributions of $q$ and the compactness parameter (see Figures \ref{type-corr1} and \ref{type-corr2}) 
suggest that the radio sources in early types may have different properties or be driven by different processes.  
However, it appears from our results that the magnitude of the spectral index across the frequency range used for  \amid\ (325 - 1400 MHz) is not sensitive to these differences.

\subsection{Relationships with Additional Data}\label{secAddData}

Relationships between spectral measurements (\ie, \amid\ and \flow) and various additional source properties are investigated in order to better understand the physical processes which shape the radio spectrum. 
Transformations are applied to the data and outliers are rejected, when necessary, to better meet the requirements of normality before testing for correlations.  For continuous variables, the significance of the relation is judged using Pearson's correlation coefficient.

\begin{figure}
\epsscale{1.20}
\plotone{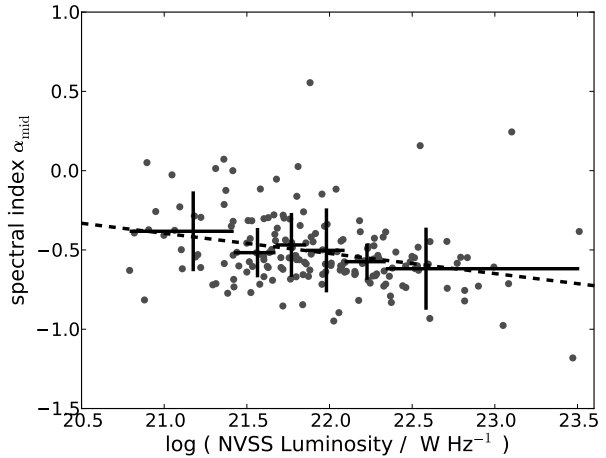}
\caption{The values of \amid\ plotted as a function of NVSS luminosity.  The crosses mark the mean and standard deviation of bins containing an equal number of sources.  A significant trend is identified whereby the higher luminosity sources in our sample are associated with steeper spectral index.  }
\label{amid-corr2}
\end{figure}

Significant trends are detected between the steepness of the spectral slope (using either \amid\ and \flow) and the source luminosity (using either NVSS or IRAS). In each case the steeper spectra are correlated with sources having higher luminosity.  One such example is shown in Figure~\ref{amid-corr2}, the case of \amid\ versus the log of NVSS luminosity, in which we have only considered sources with Hubble flow distances greater than 20 Mpc and luminosities less than $\log L = 23.6$, for $L$ in [W Hz$^{-1}$].  
The equation of the regression line determined from these data is  \amid\ $ = 2.27 - 0.13 \log L$.

\begin{figure}
\epsscale{1.20}
\plotone{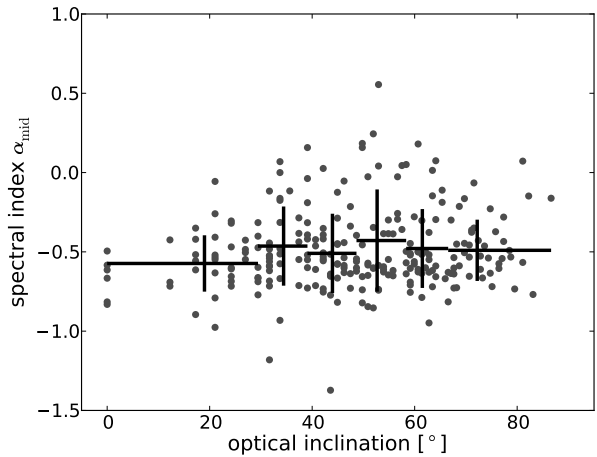}
\caption{The values of \amid\ plotted as a function of optical inclination angle and crosses marking the mean and standard deviation of bins containing  equal numbers of sources. 
We do not see this relation in our data after testing \amid,  \flow\ and projected linear size versus inclination angle, where the linear size was calculated using the optical size, inclination and Hubble flow distance. }
\label{amid-corr1}
\end{figure}

One particular relation which has been discussed in previous studies is between the low frequency spectral index and optical axial ratio (as a proxy for galaxy inclination),  motivated by observation as well as geometric models of thermal absorption.  We test our data for relationships between the optical inclination, as derived from the optical axial ratio, and spectral properties \amid\ and \flow.  
In pursuit of a better parameter to gauge the amount of expected thermal absorption, we quantify the projected path length through the galaxy disk by using our knowledge of the optical size, inclination, Hubble flow distance, and assuming disk scale height equal to 10\% of the diameter. We also test both the full galaxy sample as well as a subset containing only Hubble types Sa to Sc, since this subset is expected to better conform to our disk model.  We find no  significant relationships for any of these combinations; one example, \amid\ versus inclination angle, is provided with Figure~\ref{amid-corr1}.

\begin{figure}
\epsscale{1.20}
\plotone{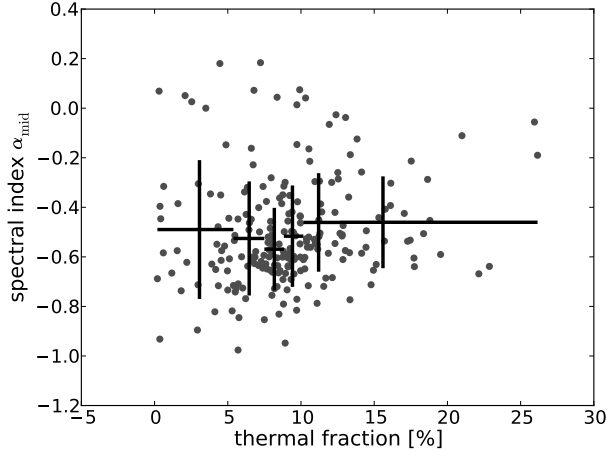}
\caption{This scatter plot is used to test for a correlation between the thermal fraction, as determined from the IRAS and NVSS fluxes (Equation~\ref{eqn-F-q}), and spectral index \amid, with crosses showing the mean and standard deviation for bins containing equal numbers of sources. No significant relationship is detected between these quantities. }
\label{amid-corr3}
\end{figure}

A final test of considerable interest is \amid\ versus $^{\mathrm{IR}}$\fthermL\ (See Figure \ref{amid-corr3})
From the simple model of power-law thermal and non-thermal emission, \eg, Figure~\ref{figToon1}, a flatter spectral index is expected to correlate with increasing thermal fraction, but no such relationship is detected.  This indicates that the simple model of optically-thin power-law components does not provide an adequate description of the spectral index \amid; this issue is discussed in greater detail in Section~\ref{secDisc}.  

After reviewing a large number of possible correlations between spectral properties (Table~\ref{table-data2}) and the tabulated source properties (Tables \ref{table-data3} and \ref{table-data4}) and rejecting those which could be attributed to selection effects, only the  relation between spectral index and luminosity is found to be significant.  
A number of supplemental tests are also conducted using  optical colors B-V and U-B, the ratio of the 60 and 100 \micron\  fluxes, the radio axial ratio and the difference between radio and optical position angles, but no significant trends are identified.

\section{Discussion and Conclusions}\label{secDisc}

For most galaxies in our sample, the fractional sensitivity for our set of flux measurements 
is not adequate to distinguish   the standard power-law  model from alternatives based on goodness-of-fit statistics.  For this reason, we focus not on fitting the individual galaxy spectra but instead on descriptive population statistics.  This section discusses our interpretation of (1) the mean values of the spectral index and generalized curvature, (2) the dispersion in the distribution of \amid, and (3) relationships between spectral characteristics of the radio emission and additional source properties. 

From the distributions of \alow, \amid, and \ahigh, we find that there is low frequency curvature present in the radio spectra of galaxies.  Comparing our result to the average spectra of the \citet{1982A&A...116..164G} sample, we find that we agree  asymptotically at the high frequency part of the spectrum but diverge at lower frequencies. One possible explanation is that the \citet{1982A&A...116..164G} sample suffers from selection effects--- approximately 40\% of their original volume limited sample was discarded because of non-detections at the lowest frequency. Other studies such as the one by \citet{1990ApJ...352...30I}, which extends down to 57.5 MHz, found that spectral curvature in the form of low frequency flattening is not uncommon in spiral galaxies. 
Therefore we conclude that low frequency galaxy spectra are, on average,  better modeled by a function having logarithmic curvature than by a single power law.  We find that $\Delta \alpha = -0.2$ per logarithmic frequency decade is the approximate magnitude for this curvature within our sample.

There are a number of physical mechanisms which could potentially produce this spectral curvature. 
Here, we consider (1) free-free absorption of a synchrotron power law, and (2) a curved synchrotron spectrum resulting from curvature in the underlying distribution of CREs. It appears unlikely that free-free absorption can effect much of the spectrum for this sample of normal galaxies, as this would require a plasma with a large emission measure and a large covering factor. 
Although compact HII regions have been found to emit  optically thick free-free emission at frequencies $>$ 10 GHz \citep[\eg,][]{2000A&A...358...95T,2002MNRAS.334..912M},  these are preferentially found in dense starburst galaxies  and do not cover a large fraction of the source. On the other hand, the hot ISM has a large covering factor but can not produce the required emission measures given typical sizes and densities. Therefore, we conclude that thermal absorption of power-law synchrotron emission can not be the primary mechanism which produces the observed curvature.


We also want to discuss some of the physical mechanisms which could be responsible for a curved synchrotron spectrum.  
Here, we adopt a model where cosmic rays are injected into the system with a power-law energy distribution \via\ shock acceleration, where the injection index $s\sim2$ \citep[for example, see][]{1983RPPh...46..973D}.  The synchrotron spectral index from this `young' electron distribution is  $\alpha \sim -0.5$ which is in good agreement with the spectral index of supernova remnants \citep[\eg,][]{1984MNRAS.209..449G}.  One description of the  spectral curvature  detected in the average spectrum (see \ref{figToon2}) is 
that we are observing a gradual transition from the injection index to a steeper spectrum at higher frequencies.  This spectral behavior is typically associated with radiative losses (\ie, synchrotron and inverse Compton), for which the loss rate is proportional to the CRE energy. However,  strong radiative cooling (an essential component in many R-FIR models) alone would produce a power-law spectrum with a much steeper spectral index $\alpha \sim -1.0$.  Here, we propose two scenarios under which radiative cooling could produce the observed curvature.  For one, the radiative cooling could transition from weak, at the lower frequencies, and become stronger at the higher observed frequencies.  
In this scenario, electrons radiating below about 1 GHz would need to be younger than their radiative lifetime.  
Alternatively, the curvature could be produced by the combination of strong radiative losses and another loss mechanism such as ionization or relativistic bremsstrahlung.  In this second scenario, the energy loss rates of these processes would need to be comparable for electrons radiating near 1 GHz.

Some studies have used the integrated spectrum to estimate physical parameters such as the density and magnetic field. For example,  \citet{2013ApJ...768...53Y} compares the results of several models applied to the spectrum of M82.  However, the variability of these results emphasizes that the estimated parameters depend upon the details of the models and their incorporated assumptions. 
An additional consideration which is often neglected when analyzing the total radio spectrum is the effect of inhomogeneous sources.  It is known that the physical conditions within galaxies (\eg, SFR,  density, temperature, magnetic field, interstellar radiation field) vary as a function of position, and  consequently,  CRE injection, radiation and  CRE energy loss mechanisms also vary accordingly.  
If different regions within the source produce distinctly different spectra, then the uniform `single box' models 
may not be adequate to interpret the net spectrum of these regions \citep[\eg, as discussed by][]{2000A&A...354..423L}. 
A number of observations have demonstrated the high degree of inhomogeneity in star forming systems, including spatial variations of the radio spectrum  \citep[\eg,][]{1991ApJ...369..320S, 1992A&A...256...10R, 1996MNRAS.281..301L, 2013A&A...557A.129T}. 
Additionally, a new study of the resolved radio continuum of two nearby galaxies  \citep{Marvil-thesis} demonstrates the importance of inhomogeneities in shaping the integrated spectrum. 
For these reasons, we elect not to pursue this type of quantitive interpretation of the spectral results.


The origin of the observed dispersion in \amid\ is another point of considerable interest.  On one hand, the dispersion is small enough that it would appear that the physical conditions among the majority of galaxies in our sample are remarkably similar.  Yet on the other hand, this distribution is broad enough that we can not explain it solely by measurement uncertainty alone, indicating that some galaxies have intrinsically steeper or flatter spectra than the average.  

One possible model which can address the intrinsic dispersion in \amid\ involves a  thermal fraction which varies within the sample, such that low \fthermL\ sources have spectra which resemble the underlying non-thermal spectrum ($\alpha_{\mathrm{NT}} \lesssim -0.7$) and high \fthermL\ sources have spectra which approach a purely free-free spectrum ($\alpha \geq -0.1$).  The right skew present in the distribution of \amid\ would further support this explanation.  However, we do not detect correlations between $^{\mathrm{IR}}$\fthermL\ and \amid\ or \flow\ as would be predicted by such a model. Additionally, the large thermal fractions needed to explain the typical \amid\ values near -0.55,  based on the decomposition model of Equation~\ref{eqn-specific-model}, are much higher than those found in similar studies.

Several other factors may contribute to the dispersion in \amid. 
We considered the effect of the relation between spectral index and radio luminosity, but this  only accounts for a small fraction of the \amid\ dispersion. 
Despite testing a large number of potential correlations, we were unable to identify any additional relationships with which to illuminate the nature of this dispersion.  The source's star formation history may also play an important role in determining the spectral index;  variations of the constant-injection model of synchrotron spectral aging predict that post-starburst galaxies will have significantly steeper spectra while sources whose star formation rate is increasing with time will have flatter spectra.  Although we tested for and found no relationships between \amid\ and optical colors U-B and B-V, a more accurate measure of star formation history may be necessary to reveal such a relationship if it exists.   Galaxy mass is another source property which we were unable to test for, and may affect the spectral index as discussed by \citet{1991A&A...246..323K}.

There are a small number of outliers in the \amid\ distribution (see Figure~\ref{figAmid}) which we have tried to better understand. Here, we discuss 16 outliers identified as having values of \amid\ outside the interval (-1,0). 
We find that their uncertainties are not significantly larger than the average uncertainties in the sample, and as such, their differences from the sample mean are highly significant.  We also find that these outliers are much more likely to lack IRAS identifications (38\%) when compared to the rest of the distribution (9\%).  For the outliers which have IRAS detections, their mean value of $q$ is 1.97, which, compared with the distribution of $q$ from  \citet{2001ApJ...554..803Y}, is about 3$\sigma$ away from the mean value (on the radio-loud side). Since these outliers are, on average, radio loud, it is plausible that those lacking IRAS identifications are also radio-loud (\ie. too faint to be included in the IRAS catalog).
There is also the issue of NVSS source identifications, for which we accepted approximately 1\% background radio sources not associated with the optical galaxy;  this would lead to $\sim$~8 objects in our sample which are likely to be background AGN.  Some spectral index outliers may instead be the result of AGN in our target sources themselves.   Further investigation of these spectral index outliers reveals that more than 60\% can be classified as AGN based on   catalog cross-referencing \citep[\ie,][]{2010A&A...518A..10V, 2012MNRAS.421.1569B} and examination of very high resolution radio continuum images. Moreover, we find that over 60\% of 22 outliers in the $q$ distribution can also be associated with AGN using these same techniques.

From the tested correlations with Hubble type, we find several distinguishing characteristics within our sample: the radio sources in early types appear to be more compact, have higher surface brightness and have excess radio emission as compared to infrared emission. 
Nevertheless, the mean spectral index of radio sources in our early types appears to be the same as the mean spectral index of radio sources in spiral types.  Therefore, it would appear that the magnitude of the spectral index can not be used to probe the physical nature of the radio source unless the spectral index takes on extreme values indicative of purely thermal or non-thermal processes.

One of the primary goals at the outset of this investigation was to measure relationships between source properties and spectral characteristics in order to better understand how the radio continuum spectrum is related to a source's physical conditions and star formation history. 
However, this analysis was limited by the sensitivity of current large-area radio survey data, which is not adequate to accurately measure spectral curvature for  individual objects in our sample. 
New large-area radio surveys with enhanced frequency coverage and sensitivity are  highly desirable to improve the detection statistics and extend the range over which the spectrum can be analyzed.  Additionally, improving the spatial resolution of the spectral data may provide an important probe of  the source's physical nature.  Furthermore, it may be necessary to expand the set of source properties  (\eg,  gravitational potential, ISM density, magnetic field strength, wind velocity) to identify stronger correlations.

\acknowledgments

JM was supported by a NRAO Grote Reber Doctoral Fellowship. 
The VLSSr, NVSS and GB6 surveys were conducted using instruments of the National Radio Astronomy Observatory.
The National Radio Astronomy Observatory is a facility of the National Science Foundation operated under cooperative agreement by Associated Universities, Inc.  Special thanks to Joe Lazio, Namir Kassim, Wendy Peters and Bill Cotton for status updates and early access to the VLSSr survey data.
 The WENSS project was a collaboration between the Netherlands Foundation for Research in Astronomy and the Leiden Observatory. We have made use of the WSRT on the Web Archive ({\tt http://www.astron.nl/wow}). The Westerbork Synthesis Radio Telescope is operated by the Netherlands Institute for Radio Astronomy ASTRON, with support of NWO.  
This research has made use of the NASA/IPAC Extragalactic Database (NED) which is operated by the Jet Propulsion Laboratory, California Institute of Technology, under contract with the National Aeronautics and Space Administration.  The VizieR \citep{2000A&AS..143...23O} archive server was used extensively for catalog queries ({\tt http://cdsarc.u-strasbg.fr/vizier/}). 


\begin{thebibliography}{50}
\expandafter\ifx\csname natexlab\endcsname\relax\def\natexlab#1{#1}\fi

\bibitem[{{Anantharamaiah} {et~al.}(2000){Anantharamaiah}, {Viallefond},
  {Mohan}, {Goss}, \& {Zhao}}]{2000ApJ...537..613A}
{Anantharamaiah}, K.~R., {Viallefond}, F., {Mohan}, N.~R., {Goss}, W.~M., \&
  {Zhao}, J.~H. 2000, \apj, 537, 613

\bibitem[{{Baars} {et~al.}(1977){Baars}, {Genzel}, {Pauliny-Toth}, \&
  {Witzel}}]{1977A&A....61...99B}
{Baars}, J.~W.~M., {Genzel}, R., {Pauliny-Toth}, I.~I.~K., \& {Witzel}, A.
  1977, \aap, 61, 99

\bibitem[{{Beichman} {et~al.}(1988){Beichman}, {Neugebauer}, {Habing}, {Clegg},
  \& {Chester}}]{1988iras....1.....B}
{Beichman}, C.~A., {Neugebauer}, G., {Habing}, H.~J., {Clegg}, P.~E., \&
  {Chester}, T.~J., eds. 1988, {Infrared astronomical satellite (IRAS) catalogs
  and atlases. Volume 1: Explanatory supplement}, Vol.~1

\bibitem[{{Best} \& {Heckman}(2012)}]{2012MNRAS.421.1569B}
{Best}, P.~N., \& {Heckman}, T.~M. 2012, \mnras, 421, 1569

\bibitem[{{Cohen} {et~al.}(2007){Cohen}, {Lane}, {Cotton}, {Kassim}, {Lazio},
  {Perley}, {Condon}, \& {Erickson}}]{2007AJ....134.1245C}
{Cohen}, A.~S., {Lane}, W.~M., {Cotton}, W.~D., {et~al.} 2007, \aj, 134, 1245

\bibitem[{{Condon}(1992)}]{1992ARA&A..30..575C}
{Condon}, J.~J. 1992, \araa, 30, 575

\bibitem[{{Condon} {et~al.}(1998){Condon}, {Cotton}, {Greisen}, {Yin},
  {Perley}, {Taylor}, \& {Broderick}}]{1998AJ....115.1693C}
{Condon}, J.~J., {Cotton}, W.~D., {Greisen}, E.~W., {et~al.} 1998, \aj, 115,
  1693

\bibitem[{{Condon} {et~al.}(1991){Condon}, {Frayer}, \&
  {Broderick}}]{1991AJ....101..362C}
{Condon}, J.~J., {Frayer}, D.~T., \& {Broderick}, J.~J. 1991, \aj, 101, 362

\bibitem[{{de Vaucouleurs} {et~al.}(1991){de Vaucouleurs}, {de Vaucouleurs},
  {Corwin}, {Buta}, {Paturel}, \& {Fouqu{\'e}}}]{1991rc3..book.....D}
{de Vaucouleurs}, G., {de Vaucouleurs}, A., {Corwin}, Jr., H.~G., {et~al.}
  1991, {Third Reference Catalogue of Bright Galaxies. } (Springer, New York,
  NY (USA))

\bibitem[{{Deeg} {et~al.}(1993){Deeg}, {Brinks}, {Duric}, {Klein}, \&
  {Skillman}}]{1993ApJ...410..626D}
{Deeg}, H.-J., {Brinks}, E., {Duric}, N., {Klein}, U., \& {Skillman}, E. 1993,
  \apj, 410, 626

\bibitem[{{Drury}(1983)}]{1983RPPh...46..973D}
{Drury}, L.~O. 1983, Reports on Progress in Physics, 46, 973

\bibitem[{{Duric} {et~al.}(1988){Duric}, {Bourneuf}, \&
  {Gregory}}]{1988AJ.....96...81D}
{Duric}, N., {Bourneuf}, E., \& {Gregory}, P.~C. 1988, \aj, 96, 81

\bibitem[{{Gioia} {et~al.}(1982){Gioia}, {Gregorini}, \&
  {Klein}}]{1982A&A...116..164G}
{Gioia}, I.~M., {Gregorini}, L., \& {Klein}, U. 1982, \aap, 116, 164

\bibitem[{{Gould}(1975)}]{1975ApJ...196..689G}
{Gould}, R.~J. 1975, \apj, 196, 689

\bibitem[{{Green}(1984)}]{1984MNRAS.209..449G}
{Green}, D.~A. 1984, \mnras, 209, 449

\bibitem[{{Gregory} {et~al.}(1996){Gregory}, {Scott}, {Douglas}, \&
  {Condon}}]{1996ApJS..103..427G}
{Gregory}, P.~C., {Scott}, W.~K., {Douglas}, K., \& {Condon}, J.~J. 1996,
  \apjs, 103, 427

\bibitem[{{Helou} {et~al.}(1985){Helou}, {Soifer}, \&
  {Rowan-Robinson}}]{1985ApJ...298L...7H}
{Helou}, G., {Soifer}, B.~T., \& {Rowan-Robinson}, M. 1985, \apjl, 298, L7

\bibitem[{{Helou} \& {Walker}(1988)}]{1988SSSC..C......0H}
{Helou}, G., \& {Walker}, D.~W. 1988, in NASA RP-1190, Vol. 7 (1988), 0

\bibitem[{{Hughes}(1991)}]{1991bja..book.....H}
{Hughes}, P.~A., ed. 1991, {Beams and jets in astrophysics} (Cambridge
  University Press)

\bibitem[{{Hummel}(1991)}]{1991A&A...251..442H}
{Hummel}, E. 1991, \aap, 251, 442

\bibitem[{{Israel} \& {Mahoney}(1990)}]{1990ApJ...352...30I}
{Israel}, F.~P., \& {Mahoney}, M.~J. 1990, \apj, 352, 30

\bibitem[{{Jones}(1970)}]{1970PhRvD...2.2787J}
{Jones}, F.~C. 1970, \prd, 2, 2787

\bibitem[{{Kennicutt}(1998)}]{1998ApJ...498..541K}
{Kennicutt}, Jr., R.~C. 1998, \apj, 498, 541

\bibitem[{{Klein} \& {Emerson}(1981)}]{1981A&A....94...29K}
{Klein}, U., \& {Emerson}, D.~T. 1981, \aap, 94, 29

\bibitem[{{Klein} {et~al.}(1991){Klein}, {Weiland}, \&
  {Brinks}}]{1991A&A...246..323K}
{Klein}, U., {Weiland}, H., \& {Brinks}, E. 1991, \aap, 246, 323

\bibitem[{{Lacki}(2013)}]{2013MNRAS.431.3003L}
{Lacki}, B.~C. 2013, \mnras, 431, 3003

\bibitem[{{Lane} {et~al.}(2012){Lane}, {Cotton}, {Helmboldt}, \&
  {Kassim}}]{2012RaSc...47.....L}
{Lane}, W.~M., {Cotton}, W.~D., {Helmboldt}, J.~F., \& {Kassim}, N.~E. 2012,
  Radio Science, 47

\bibitem[{{Lisenfeld} {et~al.}(1996){Lisenfeld}, {Alexander}, {Pooley}, \&
  {Wilding}}]{1996MNRAS.281..301L}
{Lisenfeld}, U., {Alexander}, P., {Pooley}, G.~G., \& {Wilding}, T. 1996,
  \mnras, 281, 301

\bibitem[{{Lisenfeld} {et~al.}(1998){Lisenfeld}, {Alexander}, {Pooley}, \&
  {Wilding}}]{1998MNRAS.300...30L}
---. 1998, \mnras, 300, 30

\bibitem[{{Lisenfeld} \& {V{\"o}lk}(2000)}]{2000A&A...354..423L}
{Lisenfeld}, U., \& {V{\"o}lk}, H.~J. 2000, \aap, 354, 423

\bibitem[{{Marvil} {et~al.}(2013, in preparation){Marvil}, {Eilek}, \&
  {Owen}}]{Marvil-thesis}
{Marvil}, J., {Eilek}, J., \& {Owen}, F. 2013, in preparation, {New Mexico Tech
  Ph.D. Dissertation}

\bibitem[{{McDonald} {et~al.}(2002){McDonald}, {Muxlow}, {Wills}, {Pedlar}, \&
  {Beswick}}]{2002MNRAS.334..912M}
{McDonald}, A.~R., {Muxlow}, T.~W.~B., {Wills}, K.~A., {Pedlar}, A., \&
  {Beswick}, R.~J. 2002, \mnras, 334, 912

\bibitem[{{Moshir} {et~al.}(1992){Moshir}, {Kopman}, \&
  {Conrow}}]{1992ifss.book.....M}
{Moshir}, M., {Kopman}, G., \& {Conrow}, T.~A.~O. 1992, {IRAS Faint Source
  Survey, Explanatory supplement version 2} (JPL D-10015 8/92 (Pasadena: JPL))

\bibitem[{{Murphy} {et~al.}(2011){Murphy}, {Condon}, {Schinnerer}, {Kennicutt},
  {Calzetti}, {Armus}, {Helou}, {Turner}, {Aniano}, {Beir{\~a}o}, {Bolatto},
  {Brandl}, {Croxall}, {Dale}, {Donovan Meyer}, {Draine}, {Engelbracht},
  {Hunt}, {Hao}, {Koda}, {Roussel}, {Skibba}, \& {Smith}}]{2011ApJ...737...67M}
{Murphy}, E.~J., {Condon}, J.~J., {Schinnerer}, E., {et~al.} 2011, \apj, 737,
  67

\bibitem[{{Niklas} {et~al.}(1997){Niklas}, {Klein}, \&
  {Wielebinski}}]{1997A&A...322...19N}
{Niklas}, S., {Klein}, U., \& {Wielebinski}, R. 1997, \aap, 322, 19

\bibitem[{{Nilson}(1973)}]{1973ugcg.book.....N}
{Nilson}, P. 1973, {Uppsala general catalogue of galaxies}

\bibitem[{{Ochsenbein} {et~al.}(2000){Ochsenbein}, {Bauer}, \&
  {Marcout}}]{2000A&AS..143...23O}
{Ochsenbein}, F., {Bauer}, P., \& {Marcout}, J. 2000, \aaps, 143, 23

\bibitem[{{Owen} \& {Morrison}(2008)}]{2008AJ....136.1889O}
{Owen}, F.~N., \& {Morrison}, G.~E. 2008, \aj, 136, 1889

\bibitem[{{Pohl} \& {Schlickeiser}(1990)}]{1990A&A...234..147P}
{Pohl}, M., \& {Schlickeiser}, R. 1990, \aap, 234, 147

\bibitem[{{Pohl} {et~al.}(1991{\natexlab{a}}){Pohl}, {Schlickeiser}, \&
  {Hummel}}]{1991A&A...250..302P}
{Pohl}, M., {Schlickeiser}, R., \& {Hummel}, E. 1991{\natexlab{a}}, \aap, 250,
  302

\bibitem[{{Pohl} {et~al.}(1991{\natexlab{b}}){Pohl}, {Schlickeiser}, \&
  {Lesch}}]{1991A&A...252..493P}
{Pohl}, M., {Schlickeiser}, R., \& {Lesch}, H. 1991{\natexlab{b}}, \aap, 252,
  493

\bibitem[{{Rengelink} {et~al.}(1997){Rengelink}, {Tang}, {de Bruyn}, {Miley},
  {Bremer}, {Roettgering}, \& {Bremer}}]{1997A&AS..124..259R}
{Rengelink}, R.~B., {Tang}, Y., {de Bruyn}, A.~G., {et~al.} 1997, \aaps, 124,
  259

\bibitem[{{Reuter} {et~al.}(1992){Reuter}, {Klein}, {Lesch}, {Wielebinski}, \&
  {Kronberg}}]{1992A&A...256...10R}
{Reuter}, H.-P., {Klein}, U., {Lesch}, H., {Wielebinski}, R., \& {Kronberg},
  P.~P. 1992, \aap, 256, 10

\bibitem[{{Rice} {et~al.}(1988){Rice}, {Lonsdale}, {Soifer}, {Neugebauer},
  {Kopan}, {Lloyd}, {de Jong}, \& {Habing}}]{1988ApJS...68...91R}
{Rice}, W., {Lonsdale}, C.~J., {Soifer}, B.~T., {et~al.} 1988, \apjs, 68, 91

\bibitem[{{Seaquist} \& {Odegard}(1991)}]{1991ApJ...369..320S}
{Seaquist}, E.~R., \& {Odegard}, N. 1991, \apj, 369, 320

\bibitem[{{Slee}(1972)}]{1972ApL....12...75S}
{Slee}, O.~B. 1972, \aplett, 12, 75

\bibitem[{{Soifer} {et~al.}(1989){Soifer}, {Boehmer}, {Neugebauer}, \&
  {Sanders}}]{1989AJ.....98..766S}
{Soifer}, B.~T., {Boehmer}, L., {Neugebauer}, G., \& {Sanders}, D.~B. 1989,
  \aj, 98, 766

\bibitem[{{Sramek}(1975)}]{1975AJ.....80..771S}
{Sramek}, R. 1975, \aj, 80, 771

\bibitem[{{Tabatabaei} {et~al.}(2013){Tabatabaei}, {Berkhuijsen}, {Frick},
  {Beck}, \& {Schinnerer}}]{2013A&A...557A.129T}
{Tabatabaei}, F.~S., {Berkhuijsen}, E.~M., {Frick}, P., {Beck}, R., \&
  {Schinnerer}, E. 2013, \aap, 557, A129

\bibitem[{{Tarchi} {et~al.}(2000){Tarchi}, {Neininger}, {Greve}, {Klein},
  {Garrington}, {Muxlow}, {Pedlar}, \& {Glendenning}}]{2000A&A...358...95T}
{Tarchi}, A., {Neininger}, N., {Greve}, A., {et~al.} 2000, \aap, 358, 95

\bibitem[{{V{\'e}ron-Cetty} \& {V{\'e}ron}(2010)}]{2010A&A...518A..10V}
{V{\'e}ron-Cetty}, M.-P., \& {V{\'e}ron}, P. 2010, \aap, 518, A10

\bibitem[{{Williams} \& {Bower}(2010)}]{2010ApJ...710.1462W}
{Williams}, P.~K.~G., \& {Bower}, G.~C. 2010, \apj, 710, 1462

\bibitem[{{Yoast-Hull} {et~al.}(2013){Yoast-Hull}, {Everett}, {Gallagher}, \&
  {Zweibel}}]{2013ApJ...768...53Y}
{Yoast-Hull}, T.~M., {Everett}, J.~E., {Gallagher}, III, J.~S., \& {Zweibel},
  E.~G. 2013, \apj, 768, 53
 

\bibitem[{{Yun} {et~al.}(2001){Yun}, {Reddy}, \&
  {Condon}}]{2001ApJ...554..803Y}
{Yun}, M.~S., {Reddy}, N.~A., \& {Condon}, J.~J. 2001, \apj, 554, 803

\end{thebibliography}

\end{document}